\documentclass[12pt,preprint]{aastex}
\shorttitle{}
\shortauthors{Nesvorn\'y et al.}
\begin{document}
\title{CAPTURE OF TROJANS BY JUMPING JUPITER}
\author{David Nesvorn\'y$^1$, David Vokrouhlick\'y$^2$, Alessandro Morbidelli$^3$}
\affil{(1) Department of Space Studies, Southwest Research Institute, 1050 Walnut St., \\Suite 300, 
Boulder, CO 80302, USA} 
\affil{(2) Institute of Astronomy, Charles University, V Hole\v{s}ovi\v{c}k\'ach 2, \\
180 00 Prague 8, Czech Republic}
\affil{(3) D\'epartement Cassiop\'ee, University of Nice, CNRS, Observatoire de la C\^ote d'Azur, \\Nice, 
06304, France}

\begin{abstract}
Jupiter Trojans are thought to be survivors of a much larger population of planetesimals that existed in 
the planetary region when planets formed. They can provide important constraints on the mass and properties 
of the planetesimal disk, and its dispersal during planet migration. 
Here we tested a possibility that 
the Trojans were captured during the early dynamical instability among the outer planets (aka the Nice 
model), when the semimajor axis of Jupiter was changing as a result of scattering encounters with an ice giant. 
The capture occurs in this model when Jupiter's orbit and its Lagrange points become radially displaced 
in a scattering event and fall into a region populated by planetesimals (that previously evolved from 
their natal transplanetary disk to $\sim$5 AU during the instability).
Our numerical simulations of the new capture model, hereafter {\it jump capture}, satisfactorily reproduce 
the orbital distribution of the Trojans and their total mass. The jump capture is potentially capable of 
explaining the observed asymmetry in the number of leading and trailing Trojans. We find that the capture 
probability is (6-$8)\times10^{-7}$ for each particle in the original transplanetary disk, implying that 
the disk contained (3-$4)\times10^7$ planetesimals with absolute magnitude $H<9$ (corresponding to 
diameter $D=80$ km for a 7\% albedo). The disk mass inferred from this work, $M_{\rm disk}\sim14$-28 
M$_{\rm Earth}$, is consistent with the mass deduced from recent dynamical simulations of the planetary 
instability.  
\end{abstract}

\section{Introduction}
Jupiter Trojans are populations of small bodies with orbits similar to that of Jupiter.
They clump around two equilibrium points of the three-body problem, known as the Lagrange $L_4$ and $L_5$ 
points, with semimajor axes $a \simeq a_{\rm Jup}$, eccentricities $e\lesssim0.15$, inclinations $i\lesssim35^\circ$, 
and $\Delta \lambda = \lambda - \lambda_{\rm Jup} \sim \pm 60^\circ$, where $\lambda$ is the mean longitude 
and index ``Jup'' denotes Jupiter. The angle $\Delta \lambda$ oscillates with a period of $\sim$150 yr and full 
libration amplitude up to $D\simeq70^\circ$ (see Marzari et al. 2002 for a review). The total population of 
Jupiter Trojans is estimated to be $\gtrsim$10\% of the main asteroid belt (e.g., Jewitt et al. 2000, Nakamura 
\& Yoshida 2008).

Jupiter Trojans are thought to have been captured from a much larger population of small bodies (planetesimals) 
that existed in the planetary region ($\sim5$-30 AU) when the giant planets formed. Previous theories suggested 
that the Trojans were captured during the early stages of Jupiter's growth (Marzari \& Scholl 1998, Fleming \& 
Hamilton 2000), by collisions (Shoemaker et al. 1989), effects of nebular gas (Yoder 1979, Peale 1993, Kary \& 
Lissauer 1995, Kortenkamp \& Hamilton 2001), etc. These theories imply that the inclination distribution 
of the Trojans should be relatively narrow with most orbits having $i<10^\circ$ (Marzari et al. 2002). 
By contrast, observations show a wide inclination distribution of Jupiter Trojans with inclinations up to 
$\simeq 35^\circ$. Attempts to explain the large inclinations of Trojans by exciting orbits after capture have 
been unsuccessful, because passing secular resonances and other dynamical effects (e.g., Gomes 1998, Petit et al. 
1999, Marzari \& Scholl 2000) are not strong enough.


Morbidelli et al. (2005, hereafter M05) proposed that Jupiter Trojans were trapped in orbits at $L_4$ and $L_5$ by 
{\it chaotic capture}. Chaotic capture occurs when Jupiter and Saturn pass, during their orbital migration, near 
the mutual 2:1 mean motion resonance (MMR), where the period ratio $P_{\rm Sat}/P_{\rm Jup} = 2$ (today this ratio 
is 2.49). The angle $\lambda_{\rm Jup}-2\lambda_{\rm Sat}-\varpi$, where $\varpi$ is the perihelion longitude of 
either Jupiter or Saturn, can then resonate with $\Delta \lambda$, creating widespread chaos around $L_4$ and $L_5$. 
Small bodies scattered by planets into the neighborhood of Jupiter's orbit can chaotically wander near $L_4$ 
and $L_5$, where they are permanently trapped once $P_{\rm Sat}/P_{\rm Jup}$ moves away from 2. 

A natural consequence of chaotic capture is that orbits fill all available space characterized by long-term 
stability. In M05, the planetesimals dynamically evolving from the transplanetary disk scatter off of the giant 
planets, acquire high-inclination orbits, and remain on these orbits after capture. This creates a wide 
inclination distribution of captured bodies, and resolves the long-standing conflict between previous formation theories and 
observations discussed above.

M05 placed chaotic capture in the context of the original Nice model (hereafter ONM; Tsiganis et al. 2005, Gomes et 
al. 2005; hereafter ONM), where migration of Jupiter and Saturn past their mutual 2:1 MMR is thought to trigger an 
instability during which Uranus and Neptune are scattered into the outer planetesimal disk. Their orbits 
subsequently stabilize and circularize near 20 and 30 AU by dynamical friction (Binney \& Tremaine 1987). 
There are many things to like about the Nice model, but that does not mean it is correct. It needs to be 
continually tested against all possible constraints. Indeed, several inconsistencies of the ONM have been 
already pointed out leading to model's revisions (Morbidelli et al. 2007, Levison et al. 2011).

It is now thought that Jupiter and Saturn have {\it not} smoothly migrated over the 2:1~MMR. Instead, $P_{\rm Sat}/P_{\rm Jup}$ 
probably `jumped' from $<$2 to $>$2.3 when Jupiter (and Saturn) scattered off of the ice giants 
(Uranus, Neptune or a similar-mass planet). This model, known as the jumping-Jupiter model, is required to explain 
the secular architecture of the outer planets, orbital distribution of asteroids, and dynamical survival of the 
terrestrial planets (Morbidelli et al. 2009b, 2010,
Brasser et al. 2009, Minton \& Malhotra 2009, Walsh \& Morbidelli 2011, Agnor \& Lin 2012). Moreover, encounters 
of Jupiter with one of the ice giants are required for capture of irregular satellites around Jupiter (Nesvorn\'y 
et al. 2007). 

M05's chaotic capture does not work in the jumping-Jupiter model, because the resonances discussed in M05 do not 
occur, and alternative capture models have not been investigated so far. Here we test how Jupiter Trojans can be 
captured in the jumping-Jupiter model. We find that the Trojans are most likely captured immediately after a close encounter 
of Jupiter with an ice giant. As a result of the encounter, $a_{\rm Jup}$ changes, sometimes by as much as $\sim$0.2 
AU in a single jump. This radially displaces Jupiter's $L_4$ and $L_5$, releases the existing Trojans, and can lead to 
capture of new bodies that happen to have semimajor axes similar to $a_{\rm Jup}$ when the jump occurs. We call this 
{\it jump capture}. 
\section{Capture Simulations}
We take advantage of the results published in Nesvorn\'y \& Morbidelli (2012; hereafter NM12). NM12 
performed nearly $10^4$ numerical integrations of the early Solar System's instability. The integrations 
started at the time when the giant planets were already fully formed and nebular gas was dispersed 
(presumably $\sim$3-10 Myr after the birth of the Sun; Haisch et al. 2001, Williams \& Cieza 2011). At this time, 
the giant planets were assumed to have orbits in mutual MMRs (that have been established during the 
previous stage of convergent planetary migration in the gas disk; Masset \& Snellgrove 2001, Pierens \& Nelson 2008, 
Pierens \& Raymond 2011). A disk of planetesimals was placed beyond the outermost ice giant (hereafter 
transplanetary disk). The dynamical evolution of the planets and planetesimals was then tracked,
using an $N$-body integrator, through and 100-Myr past the instability. 

\begin{figure}
\epsscale{0.5}
\plotone{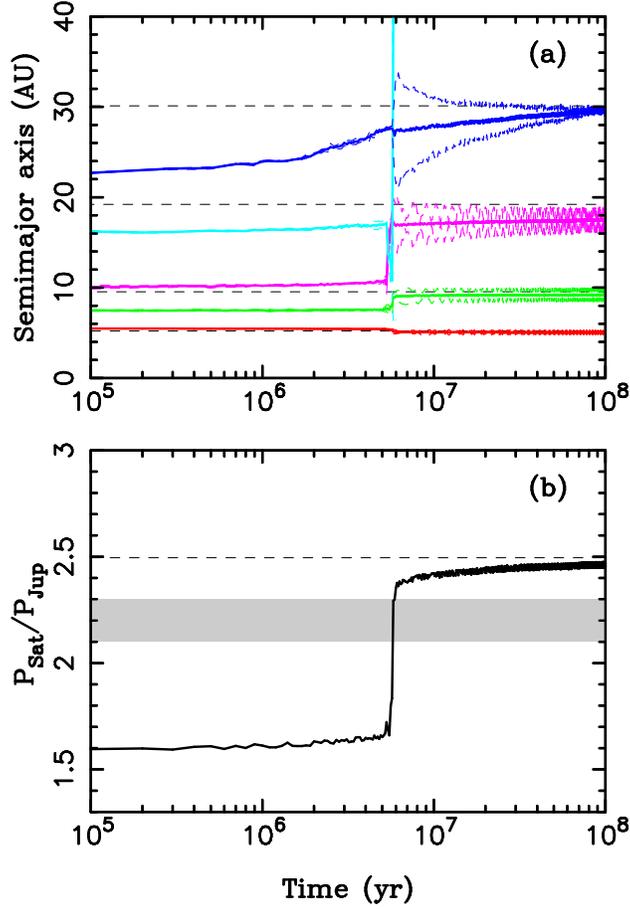}
\caption{Orbital histories of the outer planets in Case 1. The planets were started in the (3:2,3:2,2:1,3:2) resonant chain, 
and $M_{\rm disk}=20$ M$_{\rm Earth}$. (a) The semimajor axes (solid lines), and perihelion and aphelion distances (dashed 
lines) of each planet's orbit.  The black dashed lines show the semimajor axes of planets in the present Solar System. (b) 
The period ratio $P_{\rm Sat}/P_{\rm Jup}$. The dashed line shows $P_{\rm Sat}/P_{\rm Jup}=2.49$, corresponding to the period 
ratio in the present Solar System. The shaded area approximately denotes the zone where the secular resonances with the 
terrestrial planets occur.}
\label{job2}
\end{figure}

\begin{figure}
\epsscale{0.5}
\plotone{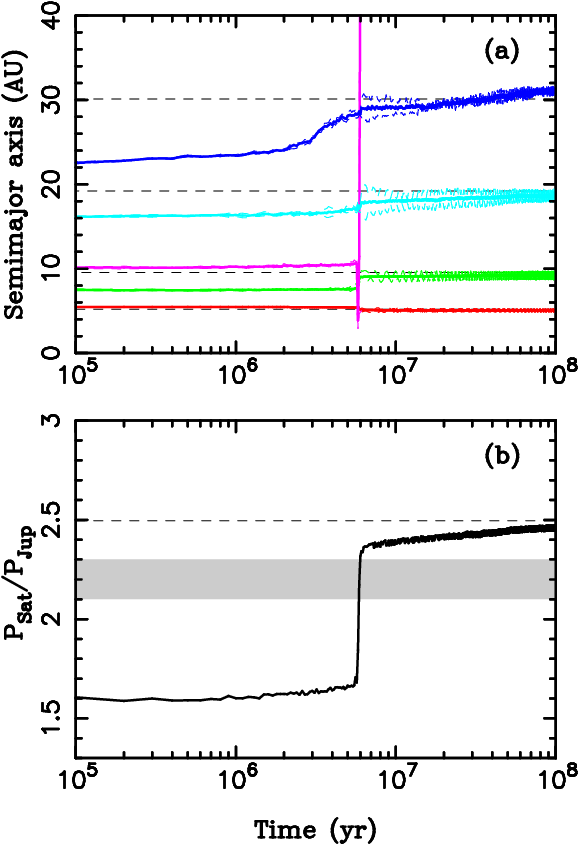}
\caption{Orbital histories of the outer planets in Case 2. See the caption of Fig. \ref{job2} for the description of orbital 
parameters shown here.}
\label{job3}
\end{figure}

\begin{figure}
\epsscale{0.5}
\plotone{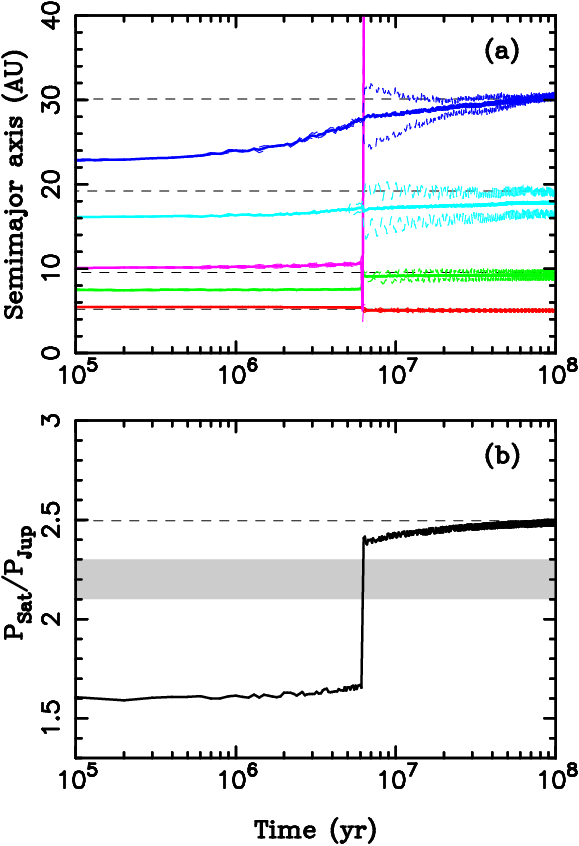}
\caption{Orbital histories of the giant planets in Case 3. See the caption of Fig. \ref{job2} for the description of orbital 
parameters shown here.}
\label{job4}
\end{figure}

Here we select three cases from NM12. Their properties are illustrated in Figs. \ref{job2}-\ref{job4}. 
In all three cases, the Solar System was assumed to have five planets initially (Jupiter, Saturn and three ice 
giants). This is because NM12 showed that various constraints (such as the final orbits of outer planets, 
survival of the terrestrial planets, etc.) can most easily be satisfied when the system starts with five initial 
planets and one ice giant is ejected during the instability (Nesvorn\'y 2011, Batygin et al. 2012).  
The case with four initial planets requires a massive planetesimal disk to avoid losing a planet, but the 
massive disk also tends to produce excessive dynamical damping and long-range residual migration of Jupiter and 
Saturn that violate constraints (Nesvorn\'y 2011, Batygin et al. 2012). It is therefore difficult to obtain a plausible 
planet evolution starting with four planets (NM12 failed to identify any in their 2670 four-planet trials). 

A shared property of the selected runs is that Jupiter and Saturn undergo a series of planetary encounters with 
the ejected ice giant. As a result of these encounters, the semimajor axes of Jupiter and Saturn evolve in discrete 
steps. While the semimajor axis can an decrease or increase during one encounter, depending on the encounter geometry, 
the general trend is such that Jupiter moves inward, i.e. to shorter periods (by scattering ice giant outward),
and Saturn moves outward, i.e., to longer periods (by scattering ice giant inward). This process leads to just the 
right kind of $P_{\rm Sat}/P_{\rm Jup}$ evolution during the instability (jumping Jupiter; see \S1).

Note that while having the fifth planet is a convenient way to obtain jumping Jupiter, the fifth planet is not 
(in itself) important for capture of Jupiter Trojans. Instead, their capture is controlled by the evolution
of Jupiter's (and Saturn's) orbit (see \S3.1). Therefore, if the future studies will identify plausible jumping-Jupiter 
cases with four planets, it is expected that the capture process described here will apply to those cases as well 
(unless the dynamical history of Jupiter's orbit will strongly differ from the one studied here).
 
NM12's simulations were performed using the symplectic integrator known as {\tt SyMBA} (Duncan et al. 1998). 
The planetesimal disk was resolved by up to 10,000 disk particles in NM12, which was sufficient for instability 
calculations, but will be insufficient here where the expected capture probability is $<10^{-5}$ (M05). To deal 
with this issue, we developed a new method that allows us to track the planetary evolution taken from the 
original {\tt SyMBA} run, and include a very large number of disk particles whose orbits are numerically integrated 
by the {\tt swift\_rmvs3} code, part of the {\it Swift} package (Levison \& Duncan 1994). This works as follows.  

We first repeat the selected NM12 jobs using {\tt SyMBA}, and record the planetary orbits at 1-yr time intervals. 
Our modified version of {\tt swift\_rmvs3} then reads the planetary orbits from a file, and interpolates them to 
any required time sub-sampling (generally 0.25 yr, which is the integration time step used here in {\tt swift\_rmvs3}). 
The interpolation is done in Cartesian coordinates. First, the planets are forward propagated on the ideal Keplerian 
orbits starting from the positions and velocities recorded by {\tt SyMBA} at the beginning of each 1-yr interval. 
Second, the {\tt SyMBA} position and velocities at the end of each 1-yr interval are propagated backward (again on 
the ideal Keplerian orbits). We then calculate a weighted mean of these two Keplerian trajectories for each planet so 
that progressively more (less) weight is given to the backward (forward) trajectory as time approaches the end of 
the 1-yr interval. We verified that this interpolation method produces insignificant errors. 

The {\tt swift\_rmvs3} jobs were launched on different CPUs, with each CPU computing the orbital evolution of a 
large number of disk particles ($N_{\rm disk}$). The initial orbital distribution of each particle set was chosen
to respect the initial distribution in the original simulation, but differed in details (e.g., the initial mean longitudes of 
particles were random), so that each set behaved like an independent statistical sample. This allowed us to build
up good statistics. To further improve the statistics, particles were cloned upon first reaching the heliocentric 
distance $r < 8$ AU. Particles reaching $r < 8$ AU were cloned by adding $N_{\rm clo}$ particles with orbits 
produced by a small (random) perturbation of the orbital velocity vector (relative magnitude $\sim10^{-5}$). 
 
We used a modest number of clones in the initial runs ($N_{\rm clo}=2$-4). Upon convincing ourselves that a more 
aggressive cloning leads to correct results (e.g., captured particles come from different clones), we used 
$N_{\rm disk}=5,000$ per CPU, $N_{\rm clo}=19$ and a large number of CPUs to obtain an effective resolution 
with 50 million disk particles in Cases 1 and 2, and 25 million in Case 3 (Table 1). The whole project was 
concluded over a period of 8 months.

The numerical integrations described above were run from a few Myr before the instability to a few Myr 
after the onset of the instability (total integration timespan of 10 Myr; Phase 1). A much longer integration was difficult 
to achieve, because the interpolation method described above had large requirements on the computer memory (planetary positions 
saved at 1-yr intervals over 10 Myr represent Gigabytes of data). Moreover, while the final planetary orbits obtained 
in NM12 matched the real orbits pretty well, they differed in details. For example, $P_{\rm Sat}/P_{\rm Jup}$ 
sometimes ended up being a bit lower ($\simeq$2.46) than in the present Solar System (2.49). This difference, 
despite being small, would affect the long-term stability of Jupiter Trojans (e.g., Robutel \& Bodossian 2009). 

We therefore continued the simulations from Phase 1 using a different method. As the planets are orbitally 
decoupled by the end of Phase 1, the integrator does not need to deal with the stochastic outcomes of planetary 
encounters. Instead, the orbital evolution of planets during Phase 2 was governed by scattering encounters with 
disk particles. As a result, the planets slowly migrated toward their current semimajor axis values. This phase 
was followed using the {\tt swift\_rmvs3} code modified to include forces that mimic radial migration (inward 
for Jupiter, outward for Saturn). The migration timescale has been set to be equal to that in the original {\tt SyMBA} 
simulations ($\simeq30$ Myr e-folding timescale). The final orbits of Jupiter and Saturn were tuned so that 
$P_{\rm Sat}/P_{\rm Jup} = 2.49$ in the end. Using artificial forces we will also slowly damped planetary eccentricities and 
inclinations (Lee \& Peale 2002), in a manner that is consistent with the original evolutions.

Phase 2 simulations were run for 100 Myr. We did not follow all disk particles during Phase 2. Instead,
we identified Trojan `candidates' by selecting particles with orbits near Jupiter's $L_4$ and $L_5$ at the end 
of Phase 1. This selection was very liberal in that most of the (hundreds of) selected bodies turned out not to 
be truly stable Trojans. The non-selected particles were discarded, which allowed us to cut down the CPU cost 
of Phase 2. One downside was that given the source population of disk particles was removed, no Trojan captures 
could have occurred in our Phase-2 simulations. This should not be a problem, however, because the population of 
particles was already depleted at this stage, and planetary evolution during Phase 2 was not favorable for capture 
(no planetary encounters, no major resonance crossings, etc.).
  
To test the long-term stability of Trojans surviving at the end of Phase 2, we performed an additional numerical 
integration over 4 Gyr. This Phase-3 integration used the original {\tt swift\_rmvs3} code 
and 0.25-yr timestep. We found that the long-term stability requirements shaved off about 50\% of Trojans from the 
population that survived at the end of Phase 2. The removed particles typically had $D>60^\circ$, large $e$ and/or 
large $i$. This result is consistent with the expected stability of Jupiter Trojans (Levison et al. 1997, 
Nesvorn\'y \& Dones 2002, Robutel \& Gabern 2006).
\section{Results}
Here we discuss the results of the numerical integrations described in the previous section. The mechanism of jump capture 
is illustrated in \S3.1. We then examine the orbital distribution of captured Trojans and compare 
it with observations (\S3.2). The efficiency of jump capture and its implications for the size distribution of 
planetesimals in the transplanetary disk are discussed in \S3.3. In \S3.4, we point the possible source of asymmetry 
between the populations of $L_4$ and $L_5$ Trojans. 
\subsection{Capture Process}
\begin{figure}
\epsscale{1.0}
\plotone{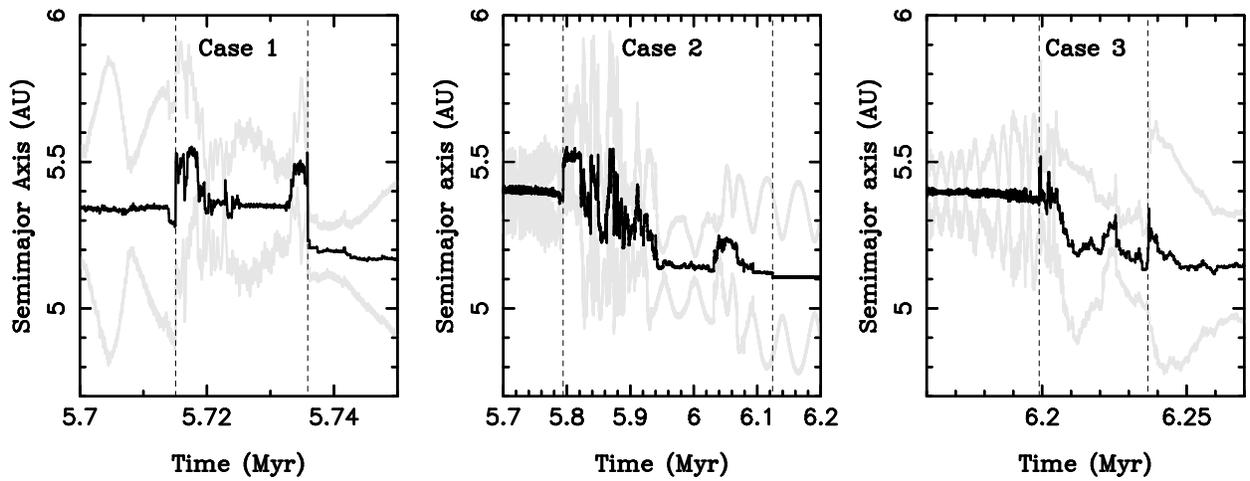}
\caption{Orbital evolution of Jupiter during planetary encounters in Cases 1 (left), 2 (middle) and 3 (right).
The solid lines show the semimajor axis (black), and perihelion and aphelion distances (gray). The dashed vertical
lines delimit the interval of Jupiter's encounters with an ice giant.}
\label{zoom}
\end{figure}

While the global evolution of planets was similar in the three selected cases (Figs. \ref{job2}-\ref{job4}), 
the detailed behavior of Jupiter's orbit differed from case to case (Fig. \ref{zoom}). This difference can be 
important for Trojan capture and is why different cases were selected in the first place. In Case 1, several
very close encounters between Jupiter and an ice giant occurred near the end of the scattering phase ($t=5.736$ Myr) 
producing a cumulative change of $a_{\rm Jup}$ from 5.53 to 5.2 AU. In Case 2, the scattering phase of Jupiter 
lasted over 300 kyr with many close encounters contributing to changes of $a_{\rm Jup}$. In contrast, Case 3 showed a 
relatively poor history of Jupiter's encounters lasting 40 kyr only.

By analyzing these different cases we found that capture of {\it most} Jupiter Trojans generally occurred during 
the stage of Jupiter's encounters. In Case 1, for example, roughly 70\% of captured Trojans that survived to 
the end of Phase 3 (hereafter the stable Trojans) started librating around $L_4$ or $L_5$ at $t=5.715$--5.736 Myr 
after the start of Phase 1. This clearly coincides in time with the period of Jupiter's encounters 
with the ice giant (see Case~1 in Fig. \ref{zoom}). By analyzing the capture histories in detail we 
found that $\simeq$50\% of stable Trojans in Case 1 were captured during the closest encounter between
Jupiter and the ice giant at $t=5.735869$ Myr, when $a_{\rm Jup}$ jumped from 5.53 to 5.3 AU (the ice giant 
was scattered to a very eccentric orbit as a result of this encounter). 

\begin{figure}
\epsscale{0.5}
\plotone{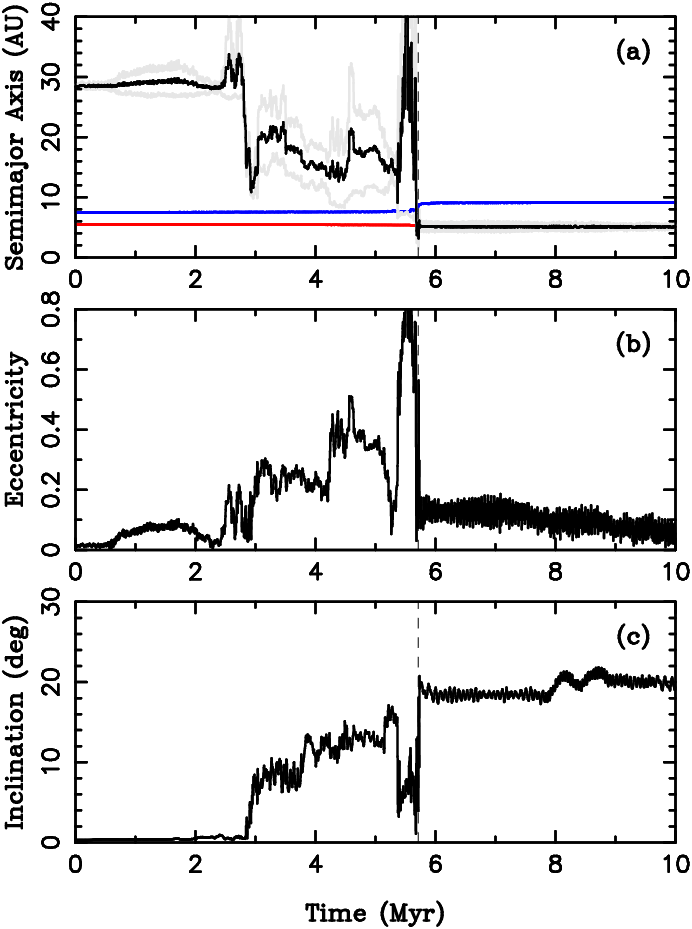}
\caption{Orbital evolution of a disk particle that was captured as a stable Trojan in Case~1: (a) semimajor axis
(black line), and perihelion and aphelion distances (gray lines) of the particle, (b) eccentricity, and (c)
inclination. Jupiter's and Saturn's semimajor axes are shown in (a) by red and blue lines, respectively.
The particle orbit remained near its starting location in the transplanetary disk for up to $t=3.5$ Myr after 
the start of the integration. The changes of $e$ and $i$ were minor during this stage. Then, at $t=3.5$ Myr, 
the particle was scattered by Neptune, evolved inward, and a series of subsequent encounters with ice giants 
raised orbit's $e$ and $i$ to moderate values. At $t=5.4$ Myr, particle's eccentricity evolved to very high values 
($e=0.8$) by encounters with Saturn. Finally, shortly before $t=5.735869$~Myr, when the closest encounter of 
Jupiter with an ice giant occurred (vertical dashed line), the particle was scattered by Jupiter. This changed its 
orbit in just the right way for capture to be possible at $t=5.735869$ Myr (i.e., $a\simeq5.3$ AU and small $e$
prior to capture). The large inclination of captured orbit was established by several scattering encounters with 
Jupiter shortly before capture.}
\label{example1}
\end{figure}

The particles captured at $t=5.735869$ Myr had special orbits just before the encounter 
($a \simeq 5.3$ AU, low $e$ and $i\lesssim30^\circ$). They were scattered to these orbits by previous encounters
with planets (Fig. \ref{example1}). They were subsequently captured in librating trajectories 
around $L_4$ or $L_5$ when the Lagrange points got displaced to $\simeq 5.3$ AU as a result Jupiter's semimajor
axis jump. This is a clear example of jump capture. In addition, roughly 10\% of the stable Trojans were jump 
captured during the previous encounter at $t=5.715$ Myr when $a_{\rm Jup}$ increased (Fig. \ref{example1}). 
They survived on librating trajectories during the closest encounter at $t=5.735869$ Myr only because they 
had the right libration phase during the encounter (so that $a \simeq 5.3$ AU).

Out of the remaining $\sim$40\% of stable Case-1 Trojans roughly 20\% were captured after the stage of 
planetary encounters was over ($t\gtrsim5.8$ Myr). Chaotic capture related to sweeping over weaker resonances, 
such as the 7:3, 12:5 and 17:7 MMRs, was responsible for these cases. The other $\simeq$20\% showed a complicated 
evolution that was difficult to classify. These cases probably correspond to jump capture during smaller jumps 
of $a_{\rm Jup}$, to chaotic capture during the irregular evolution of $a_{\rm Jup}$, or to the combination of 
both. The capture statistics in Cases 2 and 3 were broadly similar: 55\% of clear jump captures, 5\% of clear
chaotic captures, and 40\% unclear in Case 2; and 50\% of jump capture, 20\% of chaotic capture, and 30\% 
unclear in Case 3. 
\subsection{Final Orbits}
\begin{figure}
\epsscale{0.9}
\plotone{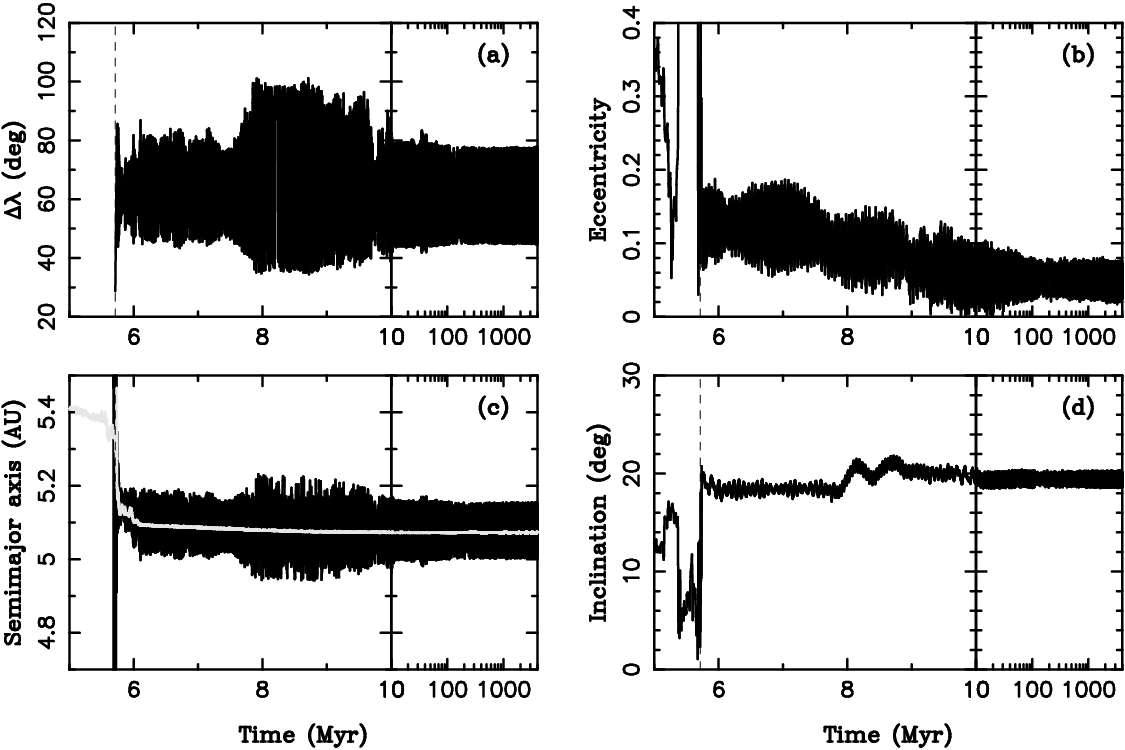}
\caption{Orbital evolution of the particle shown in Fig. \ref{example1} after its capture at $t=5.712936$~Myr 
(indicated here by the vertical dashed line). The angle $\Delta \lambda$ circulates before capture and
is not shown in panel (a) for $t<5.712936$~Myr. The bright grey line in panel (c) is the semimajor axis of Jupiter.}
\label{example2}
\end{figure}

Figure \ref{example2} shows an example of the orbital history of a stable Trojan after its capture. The changes of 
the libration amplitude seen in panel (a) are related to sweeping resonances (Robutel \& Bodossian 2009). The 
libration amplitude stabilized at $t>10$ Myr as planetary migration slowed down, and the system did not encounter 
any important resonances when $P_{\rm Sat}/P_{\rm Jup}$ approached 2.49. While the eccentricity can still significantly 
change after capture (panel b), the inclination of captured orbits typically remained nearly constant (panel d).
This shows that the inclination distribution of Jupiter Trojans is closely related to that of planetesimals near 
5 AU during the scattering phase. It is wide mainly because of scattering encounters of 
planetesimals with Jupiter and Saturn.\footnote{The wide inclination distribution of Jupiter Trojans is therefore 
unrelated to and cannot be used to constrain the inclination distribution of planetesimals in the original transplanetary 
disk. In contrast, Neptune Trojans suffered smaller inclination perturbations prior to their capture and can be 
used to this end (e.g., Nesvorn\'y \& Vokrouhlick\'y 2009).}

\begin{figure}
\epsscale{0.5}
\plotone{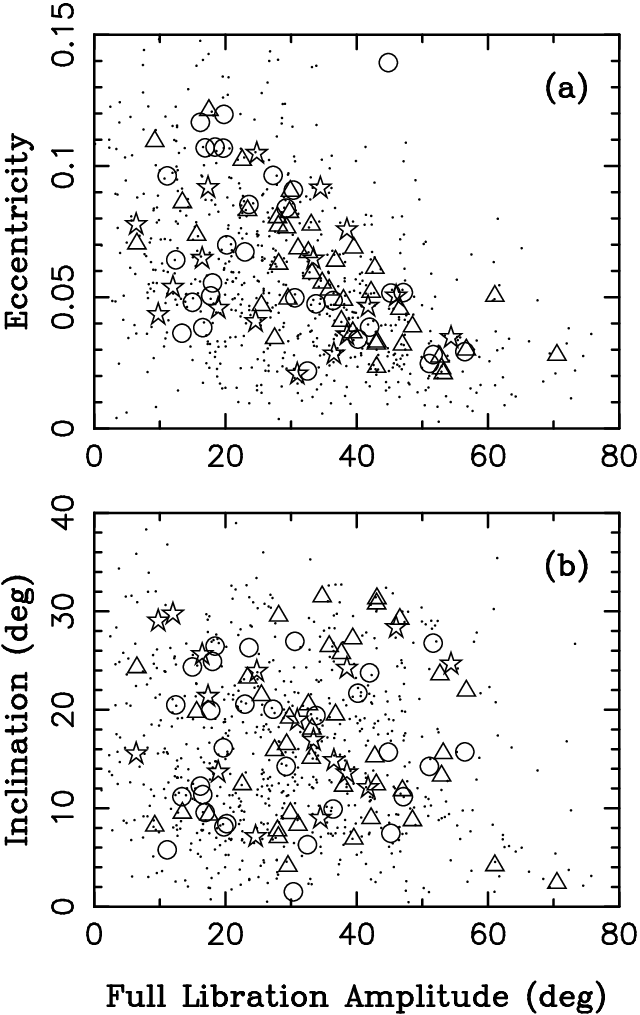}
\caption{Orbits of stable Trojans obtained in Cases 1 (circles), 2 (triangles), and 3 (stars).  
The full libration amplitude $D$ corresponds to the angular distance between extremes of 
$\lambda - \lambda_{\rm Jup}$ during libration. The black dots show the orbital distribution of real Trojans.
The proper orbital elements shown here were computed by the method described in Bro\v{z} \& Rozehnal (2011). 
We found no significant difference between the orbital distributions of $L_4$ and $L_5$ Trojans, and clumped
these distributions together.}
\label{alljobs}
\end{figure}

\begin{figure}
\epsscale{1.0}
\plotone{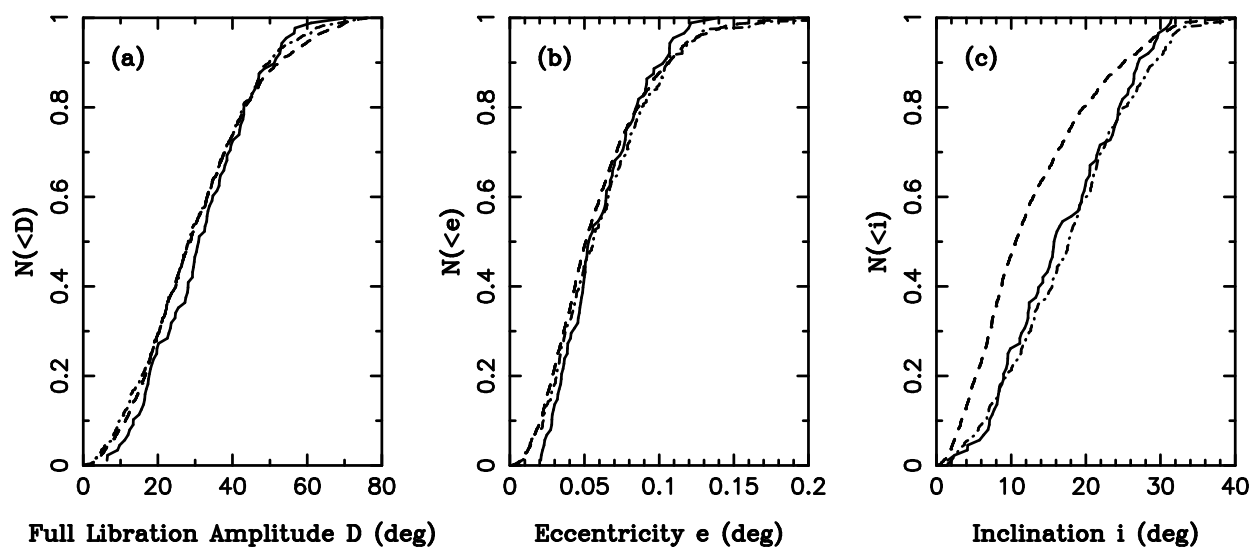}
\caption{The cumulative distribution of the (a) full libration amplitude, (b) eccentricity, and (c) inclination.
The various lines shown here denote the model distributions (solid), known Trojans (dashed), and known Trojans with 
absolute magnitude $H<12$ (dot-dashed). According to Szab\'o et al. (2007), the population of known Trojans with $H<12$ 
should be nearly complete. The difference between the dashed and dot-dashed lines in panel (c) is 
related to the incompleteness of the faint Trojans with high orbital inclinations.}
\label{ks}
\end{figure}

The orbital distribution of stable Trojans produced in our simulations very closely matches observations (Figures 
\ref{alljobs} and \ref{ks}). The distribution extends down to very small libration amplitudes, small eccentricities 
and small inclinations. These orbits are generally the most difficult to populate in any capture model. The inclination 
distribution of captured objects is wide, extending up to $i\simeq30^\circ$, just as needed.

To carefully compare the inclination distribution obtained in our model with observations, we should ideally need to 
account for the detection efficiency of objects with different inclinations. This is because most surveys look near 
the ecliptic and tend to have lower detection efficiencies for orbits with larger inclinations (e.g., Jewitt et al. 
2000). We use a magnitude cutoff to avoid this problem. According to Szab\'o et al. (2007), the Trojan population
should be (nearly) complete up to $H\simeq12$. Fig. \ref{ks}c shows an excellent agreement between our model 
inclination distribution and the one obtained with $H<12$.     

Interestingly, despite the very different histories of $a_{\rm Jup}$ in Cases 1, 2 and 3, discussed in \S3.1, 
the orbital distributions of stable Trojans that were obtained in these cases are similar (Fig. \ref{alljobs}).
This indicates that jump capture is a robust capture mechanism that is expected to produce the correct orbital 
properties of Jupiter Trojans for a wide range of jumping-Jupiter evolutions.

We applied the Kolmogorov-Smirnov (K-S) test to the orbital distributions in Fig. \ref{ks}
(Press et al. 1992). The K-S test is a statistical measure indicating whether two sets of data (the computed 
and observed orbits of Trojans in our case) are drawn from two different distribution functions, or whether 
they are consistent with a single distribution function. The K-S test gives 16\%, 68\% and 63\% probability 
that the computed and measured (with cutoff $H<12$) distributions of $D$, $e$ and $i$ are statistically 
the same, respectively. 

The lower K-S probability for $D$ is caused by a modest shortage of model orbits with $D\lesssim20^\circ$. This, 
however, varies from case to case. In Case 1, the K-S probability for $D$ is 60\%, while it is only 3\% in Case 
2. The larger probability in Case 1 may be related to the fact that most bodies were captured in Case 1 
during a single large jump of $a_{\rm Jup}$ (\S3.1). Such a clean jump can more easily produce 
$D<20^\circ$.\footnote{A two-dimensional K-S test applied to the $e$-$D$ and $i$-$D$ distributions obtained in 
Case 1 gives 25\% and 50\% probabilities, respectively, that the computed and measured distributions are the same.}
Conversely, in Case 2, a more continuous evolution of $a_{\rm Jup}$ may not allow enough bodies to evolve
to $D<20^\circ$. These conclusions will need to be checked with better statistics. 
\subsection{Capture Efficiency}
We identified 30 stable Trojans in Case 1 (out of $N_{\rm disk}=5\times10^7$ disk particles), 41 in Case 2 
($N_{\rm disk}=5\times10^7$), and 17 in Case 3 ($N_{\rm disk}=2.5\times10^7$). This corresponds to the mean 
weighted capture efficiency of $P = (7.0\pm0.7) \times 10^{-7}$ for each particle in the original 
planetesimal disk (Table 1), where the formal $1\,\sigma$ error was 
computed assuming the normal distribution.\footnote{The capture efficiency is given here for each particle in the 
original transplanetary disk. M05 instead reported, quoting, `capture efficiencies per one particle 
cycled through the system as the planets migrate through unstable Trojan configurations'. M05 found that this 
corresponds to 3.4 $M_{\rm Earth}$ in the reference simulation of Tsiganis et al. (2005). As Tsiganis et al. 
used $M_{\rm disk} \sim 35$ $M_{\rm Earth}$, the capture efficiencies reported in M05, $1.8\times10^{-5}$ to 
$2.4\times10^{-6}$, should be divided by $\sim$10 to compare them to our values.} As there are 25 known Trojans 
with $H<9$ (this sample is complete), this indicates that the planetesimal disk contained 
$\sim 25/(7\times10^{-7})=3.6\times10^7$ planetesimals with $H<9$ (corresponding to diameter $D=80$ km 
for a 7\% albedo, Grav et al. 2012). 

This is encouraging because it favorably compares with estimates obtained by other means (e.g., Nesvorn\'y et al.
2007, Levison et al. 2008, Charnoz et al. 2009, Morbidelli et al. 2009a). For example, Charnoz et al. (2009) 
suggested, using the crater record on Saturn's moon Iapetus (surveyed by the Cassini spacecraft), that the 
planetesimal disk contained $\sim$$10^7$ planetesimals with $D>80$ km. Morbidelli et al. (2009), using a synthesis
of constraints (mainly from the Kuiper belt), proposed that the disk contained $\sim$$10^8$ planetesimals with 
$H<9$. 

The stable Trojans captured in our simulations sample the full radial extent of the transplanetary disk up to 
$\sim$30 AU (at least we were not able to detect any preferential sampling based on our statistics). This shows 
that the size frequency distribution (SFD) of Trojans should be representative for the SFD in the whole 
transplanetary disk (at least) up to $\sim$30 AU (Morbidelli et al. 2009a).

M05 estimated that the present mass of the Trojan population is $M_{\rm Tro}\sim10^{-5}$ M$_{\rm Earth}$. Jewitt et al. 
(2000), on the other hand, suggested that $M_{\rm Tro}\sim 9\times10^{-5}$ M$_{\rm Earth}$. Using bulk density $\rho=1$ g 
cm$^{-3}$ (e.g., Marchis et al. 2006) instead of Jewitt's $\rho=2$ g cm$^{-3}$, and updated albedo (7\% from Grav
et al. 2012 instead of Jewitt's 4\%), we find that $M_{\rm Tro}\sim2\times10^{-5}$ M$_{\rm Earth}$. With 
$M_{\rm Tro}\sim(1$-$2)\times10^{-5}$ M$_{\rm Earth}$, we can therefore estimate that the planetesimal disk mass was 
$\sim(1$-$2)\times10^{-5}/(7\times10^{-7})=14$-28 M$_{\rm Earth}$. This is consistent with $M_{\rm disk} = 
20$~M$_{\rm Earth}$ used in NM12.   
\subsection{$L_4/L_5$ Asymmetry}
The difference in the number of leading and trailing Trojans is a long-standing problem in planetary science
(see Marzari \& Scholl 2002 for a review). This is because all capture mechanisms proposed so far, including 
chaotic capture of M05, are expected to produce symmetric distributions with the numbers of bodies in each 
swarm, $N(L_4)$ and $N(L_5)$, being nearly the same (up to statistical fluctuations). Planetary migration is 
also not expected to change $f_{45}=N(L_4)/N(L_5)$, nor is the long-term stability.\footnote{Gomes (1998) claimed that 
planetary migration can change $f_{45}$, but these changes were probably related to the choice of the initial 
orbits in Gomes (1998) rather than to an asymmetry of the effects of planetary migration itself (e.g., O'Brien 2012).} 
In addition, the existing asymmetry is not simply due to different collisional evolution of the $L_4$ and 
$L_5$ Trojan swarms, because it persists even if known collisional families at $L_4$ are removed (O'Brien \& Morbidelli 
2008).
 
Various research groups published estimates of $f_{45}$ that apply to different limiting magnitudes/sizes.
Szab\'o et al. (2007) estimated from the Sloan Digital Sky Survey (SDSS) that $f_{45}=1.6 \pm 0.1$. Survey 
with the Subaru telescope gave $f_{45} \simeq 1.8$ (Nakamura \& Yoshida 2008). Estimates from the Wide-field
Infrared Survey Explorer (WISE) gave $f_{45}=1.4\pm0.2$ (Grav et al. 2011) and 
$f_{45}\simeq 1.34$ for diameters $D>50$ km (Grav et al. 2012). Therefore, $f_{45}=1.2$-1.8 according to 
these works.

\begin{figure}
\epsscale{0.6}
\plotone{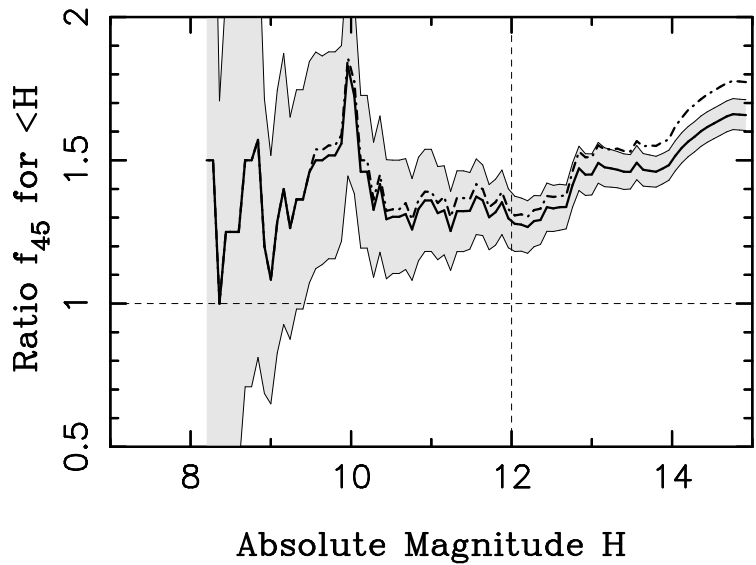}
\caption{Asymmetry between the populations of $L_4$ and $L_5$ Trojans. The solid line shows $f_{\rm 45}=N(L_4)/N(L_5)$ 
for the known Trojans as a function of absolute magnitude $H$. The Eurybates family, as identified by Bro\v{z} \& Rozehnal
(2011), was removed from $N(L_4)$ (the dash-dotted line shows $f_{\rm 45}$ with the Eurybates family included in $N(L_4)$).
The gray region denotes our $1\,\sigma$ statistical error estimate of $f_{\rm 45}$.}
\label{asym}
\end{figure}

Figure \ref{asym} shows $f_{45}$ of the known Jupiter Trojans as a function of $H$ (data from the Minor Planet Center). 
The ratio is wiggly for $H\lesssim10$, because only a very few bright Trojans exist. The statistics for $H<9$ is 
apparently not large enough to rule out $f_{45}=1$ with confidence. At the faint end, on the other hand, the 
sample is incomplete, and $f_{45}$ can be influenced by a few large collisions that generated a lot of small 
debris (e.g., Bro\v{z} \& Rozehnal 2011). 

To assess the significance of the asymmetry we find it best to use the population with $H<12$ (complete sample according 
to Szab\'o et al. 2007). The number of the known $L_4$ Trojans with $H<12$ is 361, if 9 known Eurybates family members 
with $H<12$ are removed (Bro\v{z} \& Rozehnal 2011). For comparison, there are 279 $L_4$ Trojans with $H<12$. This 
indicates that $f_{45}=1.3 \pm 0.1$, where the formal error was computed as $f_{45}[1/N(L_4)+1/N(L_5)]^{0.5}$. 
The asymmetry is therefore significant at $\simeq$$3\,\sigma$ for $H<12$.

The jump capture discussed in \S3.1 is potentially capable of producing an asymmetry. For example, the asymmetry 
can arise shortly after planetesimals are jump captured, while the ice giant still remains on a Jupiter-crossing orbit. 
During this time, the ice giant can traverse one of the Lagrange swarms and scatter captured bodies, causing the 
local population to drop. If so, the observed $f_{45}>1$ would indicate that the ice giant traversed the $L_5$ cloud 
shortly after the bulk of the Trojan population was captured. Note that this source of asymmetry does not apply to 
chaotic capture described in M05, because the orbits of ice giants do not reach down to Jupiter's orbit in M05.

From our simulations, we get $f_{45}=1.3 \pm 0.5$ in Case 1, $f_{54}\equiv 1/f_{45}=1.4 \pm 0.4$ in Case~2, and $f_{54}
=1.8\pm0.9$ in Case 3. While all three cases therefore show a formal asymmetry, the statistics is not good enough to
rule out $f_{45}=1$ at more than $1\,\sigma$. It is therefore possible that we are just seeing statistical 
fluctuations of a small sample. Unfortunately, increasing the statistics to $3\,\sigma$ with the method described 
in \S2 is not computationally feasible at this time, because we would need to increase the number of disk 
particles by a factor of $\sim$10. We leave this issue for a future work. 
\section{Conclusions}
Here we discussed a new model for capture of Jupiter Trojans. The jump capture, as we call it, occurs 
when Jupiter undergoes a series of scattering encounters with an ice giant, and $a_{\rm Jup}$ evolves 
through a number of discrete steps as a result of these encounters. We performed numerical integrations 
of jump capture, where orbits of 125 million disk planetesimals were tracked over the period of 
discrete changes of $a_{\rm Jup}$. The captured bodies were followed from the time of their capture, 
presumably some 4 Gyr ago, to the present time. The number and orbits of the surviving bodies were compared 
with observations of Jupiter Trojans. Our results are summarized as follows:
\begin{enumerate}
\item We found that the efficiency of jump capture is (6-$8)\times10^{-7}$ for each particle in the original 
transplanetary disk. This, and the number of known Trojans with $H<9$, imply that the planetesimal disk 
contained (3-$4)\times10^7$ bodies with $H<9$ (corresponding to diameter $D=80$ km for a 7\% albedo). 
The inferred mass of the planetesimal disk is $M_{\rm disk} = 14$-28 M$_{\rm Earth}$.       
\item The orbital distribution of stable Trojans obtained in our simulations provides a good match to the 
observed distribution, including orbits with small libration amplitudes, small eccentricities and 
small/large inclinations. The present wide inclination distribution of Jupiter Trojans reflects the 
distribution of planetesimals near 5 AU during the planetary instability.
\item The jump capture is potentially capable of explaining the observed asymmetry of Jupiter Trojans
($N(L_4)/N(L_5)=1.3 \pm 0.1$ for the complete sample with $H<12$). The asymmetry can be related to 
(a few significant) late passages of an ice giant near $L_5$ that presumably depleted the $L_5$ population.
Future modeling work will need to improve the capture statistics and test this possibility at a larger 
statistical confidence that it was done here.
\end{enumerate}
In a broader context, the work presented here provides support for the jumping-Jupiter model (Morbidelli 
et al. 2009b, 2010; Brasser et al. 2009), and shows a good consistency of the planetary-instability 
simulations published in NM12. 

\acknowledgments
This work was supported by NASA's Outer Planet Research programs.
Alessandro Morbidelli thanks Germany's Helmholtz Alliance for providing support through its “Planetary 
Evolution and Life” program. David Nesvorn\'y thanks the Observatoire de la C\^ote d'Azur for hospitality 
during his sabbatical year in Nice. The work of David Vokrouhlick\'y was partly supported by the 
Czech Grant Agency (grant 205/08/0064). We thank M. Bro\v{z} for helping us with the computation
of the proper elements.

\clearpage
\begin{table}[t]
\begin{center}
\begin{tabular}{lrrr}
\hline 
                               & Case 1   & Case 2  & Case 3 \\
$N_{\rm disk}$ ($10^6$)         & 50           & 50              & 25      \\
$N_{\rm cap}$                   & 30           & 41              & 17      \\   
$P_{\rm cap}$ ($10^{-7}$)       & $6.0\pm1.1$  & $8.2\pm1.3$    & $6.8\pm1.6$  \\
$f_{45}$                       & $1.3$        & $0.71$          & 0.55      \\
\hline
\end{tabular}
\end{center}
\caption{The statistics of Trojan capture. The rows are: the (1) number of disk
particles ($N_{\rm disk}$), (2) number of captured stable Trojans ($N_{\rm cap}$), (3) probability of capture
($P_{\rm cap}$), and (4) asymmetry in the number of $L_4$ and $L_5$ Trojans ($f_{45}=N(L_4)/N(L_5)$).}
\end{table}

\clearpage

\clearpage

\clearpage

\clearpage

\clearpage

\clearpage

\clearpage

\clearpage

\clearpage

\end{document}